# How Channel Segregates Originates: The Flow of Accumulated Impurity Clusters in Solidifying Steels

Dianzhong Li*, Xing-Qiu Chen, Paixian Fu, Xiaoping Ma, Hongwei Liu, Yun Chen, Yikun Luan and Yiyi Li

*Shenyang National Laboratory for Materials Science, Institute of Metal Research, Chinese Academy of Sciences, Shenyang, 110016, P. R. China*
(*Corresponding author: dzli@imr.ac.cn)

**The phenomenon, channel segregates (CS) as a result of gravity-driven flow due to density contrast occurred in the solid-liquid mushy zones[1] during solidification, often causes the severe destruction of homogeneity and even some fatal damages. Investigation on its mechanism sheds light on the understanding and controlling of the formation of solidifying metals[2,3], earth's core[4], igneous rock[5] and sea ice[6]. Until now, it still remains controversial what composes the density contrasts and, to what extent, how it affects channel segregates. Here, we show that in experimental 500kg and 100 ton commercial cast steel ingots CS originates from oxide $Al_2O_3$/MnS impurity clusters (OICs) initially nucleated from the oxide ($Al_2O_3$) particles, which induce an extra flow due to sharp density contrast between clusters and melt. The results uncover that, as OICs enrich and grow, their driven flow becomes stronger than the traditionally recognized inter-dendritic thermo-solutal convection, dominating the subsequent opening of the channels. This study extends the classical macro-segregation theory, highlights a significant technological breakthrough to control CS, and could quickly yield practical benefits to the worldwide manufacture of over 50 million tons of ingots, super-thick slab and heavy castings annually[7], as well as has general implications for the elaboration of other related natural phenomena.**

In the solidifying melt of mixture, the growing solid phase and melt usually coexist to form a typical mushy zone that is a porous-type medium[8] in nature. Among them, the interstitial melt gives rise to fluid flow (typically, natural buoyancy convection) due to the existence of a driving force induced by the density contrast between solute enriched melt and bulk liquid[9]. The most puzzling consequence of intense convection in the mushy zone, accompanied with the grain settling, is identified as so-called macro-segregation[10]. Although macro-segregation was already reported in the bronze gun barrels as early as the 1540s[11], up to the 1960s the systematic theory of macro-segregation was pioneered by Flemings and co-workers, who discovered the importance of the fluid flow in the mushy zones of solidifying alloys and derived the fundamental formula[10,12].

As one of the most typical macro-segregation defects, CS (also called "A segregates[10]", "freckle[10,13]", "chimneys[14]" or "compositional stratification[9]") has been a long-term subject in fields of metallurgy, geophysics[5,15], and geology[16], etc. Within the classical theoretical framework[10,17], CS was attributed to inter-dendritic fluid flow mainly driven by natural thermal-solutal convection. This convection is substantially stemmed from an enrichment of solutes in the inter-dendritic melt due to solute partition between solid and melt[18,19]. When the flow velocity is greater than the advancing speed of the solidification front, both local melting and flow instability occur[10,20] that result in CS in steel ingots and castings. Within these contexts, the corresponding modeling has been thus developed to simulate CS[12,21-23].

Although the classical theories of macro-segregation were extensively accepted and documented well in the textbooks, their incompleteness still exists[20]. In particular, based on those theories, the CS will unavoidably appear in huge ingots. Taking an example, for a 100 ton ingot the CS definitely occur due to the long solidification time (over 25 hours) and a slow average cooling rate (about < 5 ℃/h, see supplementary materials). Hence, the inter-dendritic fluid flow will be very intense due to the rather tall and sufficient sections. However, the examinations of our synthesized 100 ton steel ingots with a total oxygen concentration (T.O) of about $1.0 \times 10^{-3}$ wt.% revealed that the CS was not observed in the etched cross section (as discussed below). This fact does not obey the expected occurrence of CS from the classical theory, thereby motivating us to explore further mechanism for CS in heavy steel ingots.

Table 1, Operation conditions and obtained results of eight designed experiments. The AD and VCD techniques denote the aluminum deoxidation and the vacuum carbon deoxidation, respectively. T.O, C and S are the total oxygen, carbon and sulfur contents, respectively.

| Expt. | Conditions | | | Results | | | | |
| --- | --- | --- | --- | --- | --- | --- | --- | --- |
| | Mass | Poured method | Deoxi- dation | T.O ($\times 10^{-3}$ wt.%) | C (wt.%) | S (wt.%) | OICs | CS |
| **I** | 500 kg | air | AD | 5.6 | 0.47 | 0.016 | Yes | Yes |
| **II** | 500 kg | Vac. | VCD | ≤1.0 | 0.47 | 0.005 | No | No |
| **III** | 500 kg | Vac. | VCD | ≤1.5 | 0.44 | 0.013 | No* | No* |
| **IV** | 20 ton | Vac. | AD | ≤1.0 | 0.35 | 0.005 | Yes | Yes |
| **V** | 500 kg | Vac. | VCD | 2.0 | 0.07 | 0.004 | Yes | Yes |
| **VI** | 100 ton | Vac. | VCD | ≤1.0 | 0.22 | 0.005 | No | No |
| **VII** | 100 ton | Vac. | AD | ≤1.5 | 0.22 | 0.002 | Yes | Yes |
| **VIII** | 100 ton | Vac. | VCD | ≤1.3 | 0.22 | 0.003 | No | No |

Here, * denotes that both OICs and CS nearly disappear in the experiment **III**.

To address this mechanism, we first designed five experiments (**I**–**V**) as compiled in Table 1. From the cut, etched longitudinal sections of the ingots **I** and **II** (**Fig. 1a** and **2a**) the substantial differences can be observed. The experiment **I** exhibits the typical channel segregates with some narrow, vertical and center inclined, axial-symmetry strip-like chains (**Fig. 1a**), whereas in the experimental **II** the CS disappeared (**Fig. 2a**). Their difference lies in the employed deoxidation techniques, resulting in a highly distinct oxygen concentration: for the former **I** the T.O is $5.6 \times 10^{-3}$ wt.% with the AD technique, while for the latter **II** the VCD technique makes T.O below $1.0 \times 10^{-3}$ wt.% (Table 1). The further analysis reveals that the inclusions enriched in the CS region of the experiment **I** (**Fig. 1b**) are most composed of $Al_2O_3$, MnS and sometimes, minor amount of bubble-like cavities (Supplementary Materials). A special feature has been regularly observed in the CS region that MnS (**Fig. 1e**) always tends to combine together with $Al_2O_3$ (**Fig. 1b**) to form the oxide $Al_2O_3$/MnS impurity clusters (OICs), most of which have a diameter of around 5-50μm. Their typical morphologies are MnS-like impurities precipitate surrounding the centered $Al_2O_3$ (**Fig. 1c**). Of course, some main elements are also promoted to segregate, as accompanying with the occurrence of the OICs (see supplementary materials). In comparison with the experiment **I**, the amount and size of the OICs have been significantly reduced in the experiment **II** (**Fig. 2a**). These contrasting experiments between **I** and **II** imply that OICs is seem to be a crucial factor for the CS formation.

In order to further elaborate the effects of MnS and $Al_2O_3$ in OICs, in the experiment **III** keeps T.O below $1.5 \times 10^{-3}$ wt.% with the VCD technique but adjust S content to a commercial level of $1.3 \times 10^{-2}$ wt.% (**Table 1**). Similar to experiment **II**, the CS was significantly reduced (**Fig. 2b**) despite of the existence of a relatively large amount of MnS. Furthermore, although the T.O is as low as $1.0 \times 10^{-3}$ wt.% in the 20 ton crankshaft forging ingot (experiment **IV**, **Fig. 2c**) for which the S concentration is about $4.0 \times 10^{-3}$ wt.% (Table 1), the typical CS were still evidenced because local OICs were found existed due to the application of the AD technique. Even without AD technique, the CS is still robust as long as OICs exist, as elucidated by our experiment **V** (**Fig. 2d**) of a 500 kg vacuum poured ingot with a T.O of $2.0 \times 10^{-3}$ wt.% due to the VCD technique (Table 1). These facts clarify that the OICs have a primary effect on the CS formation.

Because the concentration of dissolved oxygen in steel melt is usually below $1.0 \times 10^{-3}$ wt.% before the pouring process, the $Al_2O_3$ in the CS should not mainly result from the reaction between Al and oxygen during solidification. Whereas, in the refining process, the high-melting-point oxide of $Al_2O_3$ (2030 $^oC$) is unavoidably formed, if the AD technique is adopted in traditional metallurgical practices. Amounts of $Al_2O_3$ particles with diameters below 10 μm are supposed not to float up rapidly, according to Stokes law which defines that the floating velocity is proportional to the square of the particle's diameters. Hence, these small ones will be buried into melts. During the solidification process, they enrich, slowly float up and finally involve together. Some of them flow in the mushy zones and enter the inter-dendritic regions (or zones of columnar to equiaxed transition, or in the vicinity of solidification fronts), before they directly floated up into the top melt. As an $Al_2O_3$ particle moves, it will adsorb the surrounding S, Mn or other ions to form a larger OIC (**Fig. 1c**,) which becomes more buoyant. Besides these adsorptions can be enhanced by the inter-dendritic micro-segregation, they can be chemically evidenced by our first-principles calculations. Due to the strong electronic hybridizations among S and its nearest-neighboring O and Al on the surface, α-$Al_2O_3$ favorably traps the free S ion and binding with Mn to initially nucleate MnS-like cluster, as indicated by the negative adsorption energy of about −2.0 eV in **Fig. 3b** (supplementary materials). The calculations also highlighted a trend (**Fig. 3c**), as nucleated Mn+$n$S-like clusters coarsen to crystalline phase on the surface, their interface binding energies with the substrate of $Al_2O_3$ certainly become weaker and weaker, eventually resulting in their separation (see **Fig. 1b**).

Furthermore, within the theoretical model of multiphase flow (supplementary materials), the moving trajectory of a dispersed OIC particle in steel melts has been calculated by encompassing the effects of both the melts on the particle trajectories and the particles on the melts flow distribution, while the interactions between particle and solidification were neglected. The calculated size-dependent moving velocity of a single freely floating particle (i.e., $Al_2O_3$) in the Fe−C (0.36 wt.%) melts is compiled in **Fig. 3d**. For a particle with a diameter from 10 μm to 100 μm, its floating velocity ranging from 0.043 mm/s to 4.11 mm/s is faster than the averaged advancing velocity (approx. 30 μm/s) of the solidification front. In real mushy zones the OICs can't float as freely as in simulations due to the capture or block of solidified melts. However, these results still provide a general impression that the OICs can move in the melt and transport in a wide region. As the solid particle moves, it drags together the surrounding melt which, in turn, results in the solute (*i.e.*, carbon) accumulation into the moving passage. This effect becomes more important, when lots of impurity particles (*i.e.*, oxide, MnS, *etc*) are accumulated into a larger OIC. In addition, the flow induced by OICs plays a significant role on the CS formation in steels because the solute convection, which is conventionally recognized as the most important factor that leads to the enrichment of solutes, is rather weak (**Fig. 3e**), as revealed by phase-field simulations in coupling with fluid flow dynamics (supplementary materials). Therefore, it is further convinced that the OICs flow dominates the CS formation.

This class of OICs driven flow has been unfortunately neglected in long-term scientific and engineering practices as well as in fundamental macro-segregation theory, although the effect of double oxides films in casting melts was ever recognized to nucleate air bubbles by Campbell[24]. We hence defined this neglected flow as an extra flow. Mechanically, it creates effectively many small chimneys in the mushy zones due to its large buoyant force, not only speeding up the surrounding melt to flow, but also leading to the enrichment of some low-melting point impurities (i.e., some sulfides) and bubbles. During the solidification, OICs congregate in a certain distance away from the side wall due to the temperature fields, and those OICs obstructed by dendrite arms will continue to move incline-upward. Because of the stuck or adsorbed solute elements and sulfides (i.e., MnS) around the OICs, they would re-melt or erode the dendritic trunks and branches, enrich and float up together, and eventually accumulate to macro-scale CS, as sketched in **Fig. 4**. In addition to this extra flow, it is necessary to note that in the small mushy zones between dendrites the interfacial tension driven flow may play a certain role in coagulating the impurity particles[25] and reshaping the clusters in the strip-like chains through the capillary force in streams. Also, we believe that the current analysis of CS may be suitable for V-type segregates[20,26], which normally appears in the bottom of the top feeder or in the continuous castings.

As a new mechanism for the CS formation, it has to be examined in the engineering practices. To yield this aim, three 100 ton industrial ingots (**VI**, **VII**, and **VIII** in Table

1) all with a 2.4m diameter and a 3.6m height were produced by vacuum degassing and vacuum pouring processes. As expected, there was almost no any CS observed in the whole cross section of the experimental **VI** (see, **Fig. 5a**), matching our currently theoretical analysis due to the removal of OICs. Unfortunately, from this experiment **VI** we still learned the occurrence of the typical centered-line porosity defects in the centered equiaxed grain zone (**Fig. 5d**). For the experimental **VII**, we thus adopted the preheated and insulation feeder techniques to reduce centered-line porosity defects while a little amount of Al is further fed in the refining ladle furnace to analyze the effect of OICs on CS. The results reveal the presence of several slim CS (**Fig. 5b**) in the coarsened dendritic grain zones (**Fig. 5e**) and the formation of $Al_2O_3$ with a T.O < 1.5 $\times 10^{-3}$ wt.% but no centered-line porosity defects. This experiment further validated the conclusion: OICs are a source initializing the CS. Combining the techniques of experiments **VI** and **VII** but, absolutely, avoiding using the AD technique, the third 100 ton ingot (experiment **VIII**) has been produced, successfully. This one is as perfect as we expected. There is no any CS in the body of the ingot except for an inclusion (**Fig. 5c**).

In the traditional metallurgy, the CS in steels is mainly eliminated through rapid cooling, mechanical vibration and electromagnetic stirring, etc. Increasing the cooling rate not only accelerates the freezing of liquid melt but also limits the floating of the OICs. Mechanical vibration and electromagnetic stirring externally alter liquid flow, naturally affecting the flow behavior of OICs. Therefore, to some extent, these technologies have certain effects on the prohibition of CS. However, these technologies have their intrinsic limitations to the application in heavy and tall steel ingot due to the thick thermal diffusion boundary layer and the heavy bulk melts. Most strikingly, our current findings are changing the traditional view to prevent the CS formation from those physical metallurgical techniques to the currently used chemical metallurgical purification. Namely, through purifying solidifying melts, to keep the T.O down to an extremely low level (typically, T.O <1.0 $\times 10^{-3}$ wt. %) and to jointly control the impurities (i.e., S, P, $Al_2O_3$ etc), the CS can be eliminated in engineering. This purification technology represents a significant technological breakthrough to pin the CS occurrence, being controllable and efficient for the industrial manufacture of heavy steel ingots. This is a substantially innovated way, different from the traditional ones by controlling thermal-solutal convection. The role of OICs on extra flow highlights a new horizon, showing a substantially difference from the traditionally recognized thermal-solutal convection.

Finally, the results that we have discussed demonstrate this mechanism could be used to investigate whether this conclusion also holds for other oxides (i.e., $TiO_2$ and $ZrO_2$) associated with some minor and trace elements (i.e., O, H, N, S, P, As, Sb, Sn, Pb and Bi) in ingots, super-thick slab and heavy castings as well as other related wide natural phenomena (i.e., the formation of earth's core, rivers, geological faults, magma, and sea ice)[4-6,15,16]. The indication of the buoyancy role, for example, affected by the $MgO/SiO_2$ ratio in the liquid phase was already addressed[27] in the research of earth's deep mantle.

## Methods

**Experiments:**

   A) Four commercial steel ingots (500 kg carbon steels **Experiments I, II, III** and a low alloying steel **Experiment V**) were experimentally synthesized with sand mould. 1) **Experiment I** of C 0.47, Si 0.26, Mn 0.54, S 0.016, P 0.020, T.O 0.0056 and Fe balanced in its chemical compositions (measured, wt.%). The steel was melted at 1600℃ by induction furnace, and poured at 1550℃ in the atmosphere after the Al deoxidation (AD) process. 2) **Experiment II** of C 0.47, Si 0.44, Mn 0.49, S 0.005, P 0.005, T.O < 0.001 and Fe balanced. It was refined and poured in vacuum condition with both Al-free and vacuum carbon deoxidation (VCD) techniques. 3) **Experiment III** of C 0.44, Si 0.51, Mn 0.68, S 0.013, P 0.006, T.O≤0.0015 and Fe balanced. The experimental methods are the same as the **experiment II**. 4) **Experiment V** of C 0.07, Si 1.34, Mn 2.32, Cr 9.62, W 1.52, V 0.25, S 0.005, P 0.007, T.O ≤0.0020 and Fe balanced. The experimental methods are the same as the **experiment II**. In addition, for all these four ingots, the sand mould was used to produce a 500 kg round ingot. The ingot was cut in half along the longitudinal axial. After the ingot was grinded, polished and etched by the 20% $HNO_3$-5%$H_2SO_4$-$H_2O$ solution, the 20% $HNO_3$-$H_2O$ solution, and the 5% $HNO_3$-$H_2O$ solution respectively, the channel segregates was observed. The inclusions in samples cut from the channel segregates zone were observed and identified by the scanning electron microscopy (SEM) and energy dispersive spectra (EDS).

B) **Experiment IV** (the forging crankshaft using a 20 ton ingot) of C 0.35, Si 0.34, Mn 0.74, Cr 1.46, Ni 1.58, Mo 0.19, S 0.004, P 0.005 T.O 0.0010 and Fe balanced in its chemical compositions (measured, wt.%). The ingot was produced by the electric arc furnace, ladle furnace, vacuum degassing and the AD technique as well as bottom filling in the atmosphere. The measured sample was taken from the crankshaft body after the forging and heat-treatment processes.

C) Three 100ton 30Cr2Ni4MoV ingots (**Experiments VI**, **VII**, and **VIII**) were experimentally synthesized with cast iron mould. 1) **Experiment VI** with C 0.22, Si 0.01, Mn 0.13, Cr 1.7, Ni 3.4, Mo 0.30, V 0.086, S 0.005, P 0.006, T.O ≤ 0.001 and Fe balanced in its chemical compositions (measured, wt.%). 2) **Experiment VII** of C 0.22, Si 0.07, Mn 0.06, Cr 1.6, Ni 3.4, Mo 0.28, V 0.08, S 0.002, P 0.005, T.O≤ 0.0015 and Fe balanced. 3) **Experiment VIII** of C 0.22, Si 0.01, Mn 0.06, Cr 1.67, Ni 3.6, Mo 0.27, V 0.08, S 0.003, P 0.005, T.O ≤ 0.0013 and Fe balanced. All the three ingots VI, VII and VIII, were produced by the following process: electric arc furnace, ladle furnace, vacuum degassing, mould stream degassing. The vacuum pouring temperature is controlled to 1575℃. It needs to be emphasized that for both **VI** and **VIII** vacuum carbon deoxidation technique was adopted but for **VII** AD technique was adopted and 0.014 wt.% Al was fed in ladle furnace to reduce the oxygen concentration. Afterwards, the three ingots were cut in half along the longitudinal axial. After the ingot sections were grinded, polished and etched by the 20% $HNO_3$-5%$H_2SO_4$-$H_2O$ solution, the 20% $HNO_3$-$H_2O$ solution, and the 5% $HNO_3$-$H_2O$ solution respectively, the channel segregates and microstructures were examined. The experiments demonstrated that the measured solidification times were as long as 27 hours for VI and 32 hours for VII and VIII.

**Theories and computations:**

A) **First-principles calculations:** The modeling simulations have been performed using the Vienna Ab initio Simulation Package (VASP)[28] within the framework of the density functional theory (DFT) based on plane-wave method. We adopted the generalized-gradient-approximation (GGA) within the Perdew-Burke-Ernzerhof (PBE) parameterization scheme for the exchange-correlation functional. Brillouin zone integrations were performed for the k-mesh $5 \times 5 \times 1$ according to Monkhorst and Pack technique. The energy cutoff for the plane-wave expansion of eigen functions was set to 400 eV. Optimization of structural parameters was achieved by the minimization of forces and stress tensors. The simulation mainly focuses on the interactions between S atom, Mn+$n$S atom complex and the surface of solid alumina which quite often appears in the steel melt. To establishing the model system, we first make two assumptions. They are (1) S element exists as S atom and Mn+$n$S atom complex in the melt, and (2) the interfacial interaction of alumina, S atom and Mn+$n$S atom complex with the iron melt is ignored. We selected α-alumina (α-$Al_2O_3$) as a modeling structure. The optimized lattice constants, enthalpy of formation [ΔH = −153.99 kJ/(mole of atoms)], bulk band gap ($E_g$ = 5.9 eV at the Γ-point) of bulk α-$Al_2O_3$ are in agreement with the previously reported results. Along the [0001] direction, the stacking sequence can be viewed as ….O-Al-Al-O-Al-Al-O…. The distance between any two adjacent O layers is about 2.2 Å in its bulk equilibrium phase. Thus, three (0001) surface terminations are available, namely, Al-, AlO- and O-terminated surfaces. Consistent with the reported results, the clean AlO-terminated (0001) surface is energetically most favorable, with the lowest surface energy of 1.68 J/m$^2$. Therefore, all our calculations are based on the non-dipole

AlO-terminated surface, which denotes a single outmost Al-layer plus a sub-outmost O-layer. The optimized surface structure indicates that the space distance between the topmost layer and the second layer is significantly reduced ($\Delta l \approx 0.738$ Å being about 78% of the unrelaxed distance. This results show that the topmost Al atom is now almost in the same layer with the second oxygen layer. This situation is mainly due to that the coordination number of the topmost Al atom is much less, as compared in the crystal. In order to simulate the adsorption on the surface of α-$Al_2O_3$, we built the 18-layer-thickness slab with a 15 Å vacuum depth along the [0001] direction, together with the $2 \times 2$ dimension surface unit cell. For all surface calculations, the bottom nine layers are always kept frozen and the other nine layers are allowed to be relaxed.

B) **Macro/meso-scale simulations:** The motion of OICs in steel melt is calculated using the discrete phase model in the computational fluid dynamics software Fluent 6.2.6[29]. In the discrete phase model, trajectories of particles moving in the flow field are computed using a Lagrangian approach. The floating velocity varies with the size of a particle through the drag force in the acceleration equation of particle. Details of the Lagrangian modelling approach for the treatment of particle motion have been reported elsewhere[30]. The "Two-way Coupling" method in which fluid phase influences particulate phase via drag and turbulence and particulate phase impacts fluid phase via source terms of mass, and momentum, is adopted to predict the floating trajectory of one OIC in steel melt. In calculations, it is assumed that a particle is spherical and floats up freely from the bottom in a bulk melt without any interaction with other particles and solidification. Before the moving of the particle, the liquid is stationary. The diameter of the injected particle is set in the region of 10 to 100 μm which includes the size of the impurity cluster usually observed in the channel segregates. The density of the molten carbon steel Fe–0.36 wt.% C, $\rho_l$, and $Al_2O_3$ particle, $\rho_p$, are 6.99 g/cm$^3$ and 3.64 g/cm$^3$, respectively. The dynamic viscosity of the steel melt, $\mu$, is equal to 0.0042 Pa s.

To be aware of the solute convection ahead of the dendritic solidification front, the quantitative two-dimensional (2D) Karma phase-field model for binary alloy solidification[31] with coupling of fluid flow dynamics is adopted to investigate the evolution of the solid-liquid interface under condition of melt flow. In this model description of crystallization, the evolution of phase state is described as a function of an order parameter $\psi$, composition $C$ and temperature $T$. The order parameter $\psi$ has a constant value in solid and liquid phases and varies smoothly across the thin diffuse solid-liquid interface. Transport of solute in liquid is not only by diffusion, but also by convection. Assuming the fluid is incompressible, the Navier–Stokes equations are adopted to describe the fluid motion. The gravity-driven natural convection induced by solute and thermal expansion are taken into account via a source term in the viscous equation using a Boussinesq approximation[1]. In the simulations, to clearly understand the evolution of thermal-solutal convection in the regions between columnar dendrites, a conventional upward directional solidification of Fe–0.36 wt.% C alloy was employed with different processing parameters. The calculations were performed on a 2117 μm $\times$ 6880 μm rectangular grid through an adaptive mesh and parallel computing procedure. The vertical temperature gradient was set to be 37 °C/cm, in which the top is hotter than bottom, and the pulling speed varied from 13.5 to 1081.1 μm/s. More details of the model and computational parameters used in the numerical simulations can be seen in the supplementary materials.

**Acknowledgements:** The authors acknowledge the valuable discussions with Prof. J. Campbell from the University of Birmingham, U.K. and Dr. V. M. Shcheglov from Physico-Technological Institute of Metals and Alloys, National Academy of Sciences of Ukraine, the experimental help with Xiuying Gai and Jingping Cui from the Technique Support Division, the contributions from Lijun Xia, Xiuhong Kang, Mingyue Sun, Xiaoqing Hu, Dongrong Liu, Baoguang Sang, Pengyue Wei, Yanfei Cao from the IMR, as well as the financial support from the international cooperation project (No. 2010DFR70640) from the Ministry of Science and Technology, China.

**Author Contributions:** D. Z. L. proposed the effect of OICs on channel segregates. P. X. F., H. W. L. and Y. K. L. performed the ingot casting, compositions analysis and channel segregation observation. X. P. M. performed the inclusion measurements and analysis in channel segregates regions. X.-Q. C. performed the density functional theory calculation and Y. C. performed macro/meso-scale simulations. Y. Y. L. proposed the original problem and supervised the investigation. D. Z. L. and X.-Q. C. wrote the paper with the preparations from all other authors. All authors contributed to the discussions of the manuscript.

**Author Information**: Correspondence and requests for materials should be addressed to D. Z. Li (dzli@imr.ac.cn).

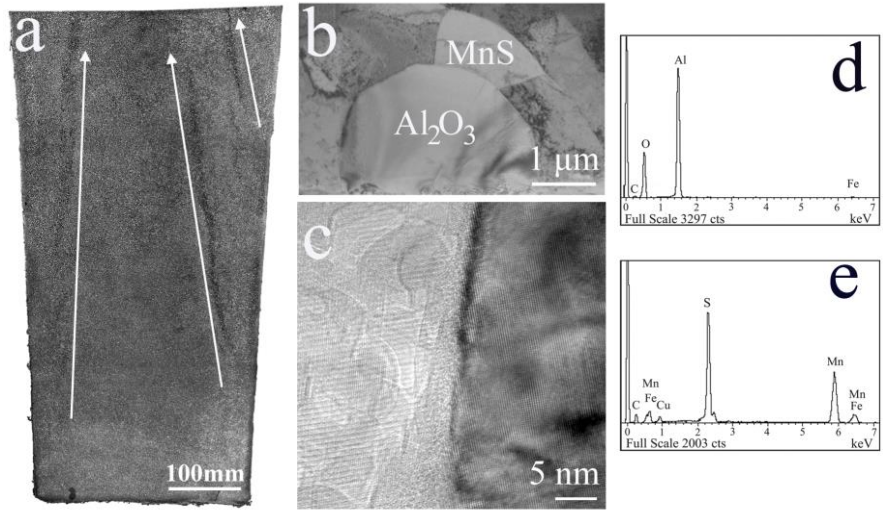

**Figure 1 Occurrences of CS and the OICs. a**, **Experiment I**: The sectioned surface of Al-deoxidation 500 kg ingot cut along the axle plane. The surface was etched by the dilute nitric acid to display the microstructures and channel segregates as marked by arrows. **b**, TEM image showing the MnS combines together with the $Al_2O_3$ to form OICs in the region of channel segregates. **c**, High-resolution TEM image to analyze the interface between MnS and $Al_2O_3$. **d** and **e,** The EDS results for $Al_2O_3$ and MnS in **b**.

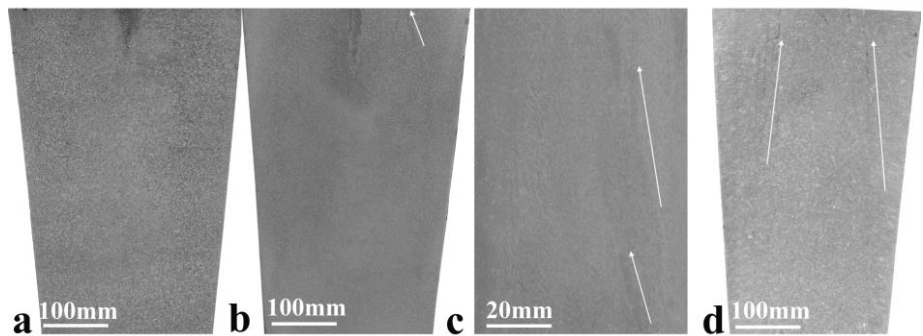

**Figure 2 Existences and eliminations of CS.** The sectioned surfaces of three vacuum poured 500 kg ingots cut along the axle plane (**a, b, d**) and of a crankshaft forging using 20ton ingot (**c**). **a**, **Experiment II**: channel segregates disappears with VCD technique to T.O $\leq 1.0 \times 10^{-3}$ wt.%; **b**, **Experiment III**: channel segregates was significantly reduced with VCD technique to T.O $< 1.5 \times 10^{-3}$ wt.%; **c**, **Experiment IV**: channel segregates appears with AD technique for 20 ton ingot due to the OICs presence even T.O is as long as $1.0 \times 10^{-3}$ wt.%; **d**, **Experiment V**: channel segregates still occurs with VCD technique to $2.0 \times 10^{-3}$ wt.% with the presence of a little amount of OICs. The arrows mark the presence of channel segregates.

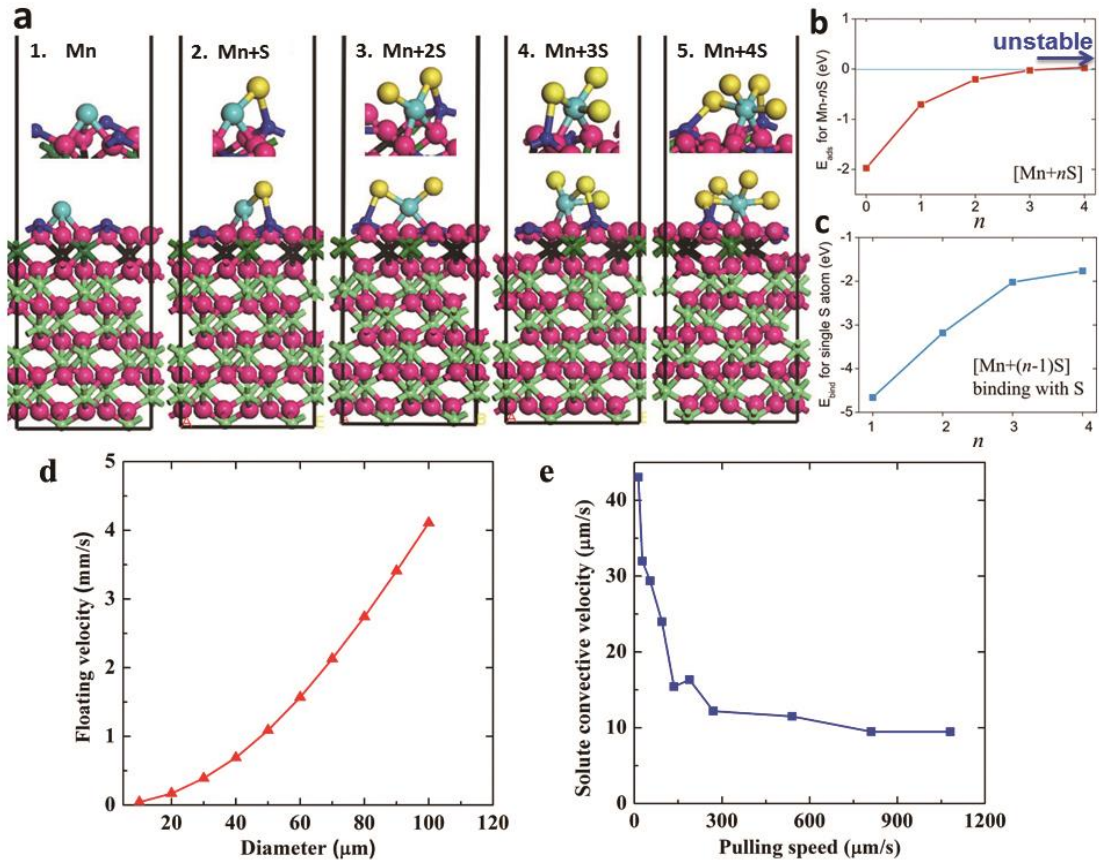

**Figure 3 Adsorption of Mn+$n$S clusters on the Al$_2$O$_3$ (0001) surface to nucleate OICs and their driven extra-flow. a**, The first-principles simulated nucleating process Mn+$n$S clusters by trapping S and Mn ions on the Al$_2$O$_3$ surface. The upper panels denote the local geometric structural details for the Mn or Mn+$n$S adsorptions. Blue and yellow balls denote Mn and S ions, respectively. It has been noted that, depending on the S introduction to the Mn+$n$S atom complex the bonding length between Mn and the nearest neighboring O1 atoms first slightly reduced from $n=0$ to $n=2$ and then it increases for the cases of $n=3$ and 4. This feature corresponds to the stable adsorption for $n=0$, 1 and 2 due to the enhanced attraction between the trapped Mn and O1 atoms on the surface. The increasing distance for $n=3$ and $n=4$ cases reveal the weakening of the adsorption. **b**, The adsorption energies of Mn+$n$S atom complex on the surface as a function of the trapped S ions. **c**, The binding energy obtained from the Mn+$n$S complex with respect to the extra-S atom and the already formed Mn+$(n-1)$S complex on the Al$_2$O$_3$ (0001) surface. **d,** Calculated floating velocities of Al$_2$O$_3$ particles vary with the particle sizes in a bulk molten steel. **e,** The phase-field simulated of the velocity of convective flow driven by solute ahead of the columnar dendritic solidification front in directional solidification of 0.36 wt.% C steel with various pulling speeds at a fixed thermal gradient $G = 37$ ℃/cm.

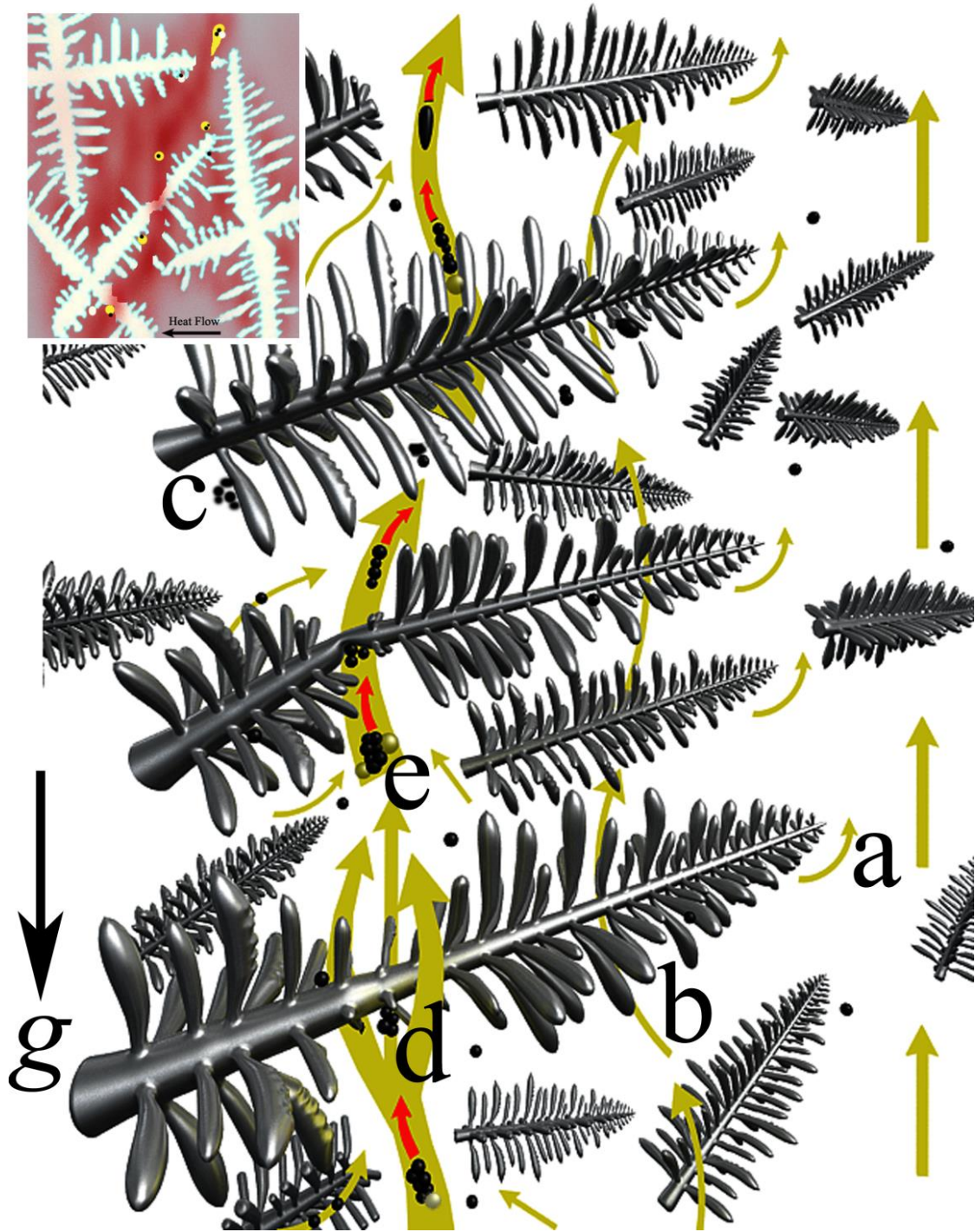

**Figure 4 The three-dimensional schematic of flows among inter-dendritic regions around the sidewalls of ingot.** Inset: two-dimensional project of inter-dendritic regions. Five flow modes are highly suggested: **a**, Solute-enriched melts run along the primary dendrite into the bulk melt and further float up in front of the dendrite tips; **b**, Solute-enriched melts vertically flow up across the dendrite trunks and branches; **c**, The floating of OICs is blocked by the dendrites; **d**, Together with the floating of accumulated OICs and minor amounts of bubbles, further re-melt and erode the dendrites to form an extra flow. The flow

channels are further formed by the extra-flow and the introduction of its correlated solute-enriched melts; **e**, The interfacial tension driven flow coagulates and reshapes OICs to strip-like chains. Solid spheres and yellow balls denote impurity particles and gas bubbles, respectively. Here, g denotes the gravity.

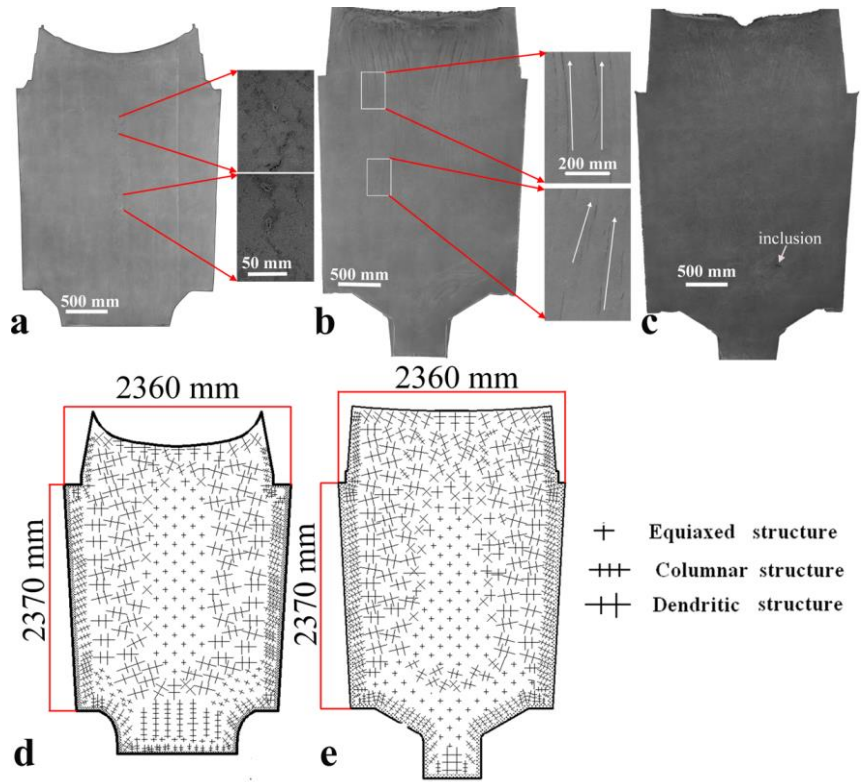

**Figure 5 The sectioned and etched surfaces and the sketches of the solidification microstructures of 100 ton 30Cr2Ni4MoV ingots with a 2.4 m diameter and a 3.6 m height.** **a**, Experiment VI: with VCD technique no channel segregates and the magnified centerline shrinkage porosities are visualized. **b**, Experiment VII: VCD technique and a little amount of Al were used for deoxidation. The local slim channel segregates are marked by arrows. **c**, Experiment VIII: with VCD technique, no channel segregates has been observed in the ingot body, except for an inclusion as marked by an arrow. **d** and **e** are the sketches of the solidification microstructures from the sectioned and etched surfaces of the ingots VI and VII, respectively. They show the typical grain zones from the first surface fine equiaxed, the second long and dense columnar, the third coarse dendritic to the fourth centered coarse equiaxed zones. Noted that the slim channel segregates in the VII ingot appears in the third zone and its connected part of the fourth zone.